\documentclass[times, 10pt,twocolumn]{article}
\usepackage{latex8}
\usepackage{times}

\usepackage{epsfig}
\usepackage{amsmath}
\usepackage{setspace}
\usepackage[small]{caption}

\title{\vspace*{-0.5in}Interconnection Networks for Scalable Quantum Computers\vspace*{-0.15in}}

\author{Nemanja Isailovic, Yatish Patel, Mark Whitney, John Kubiatowicz \\
Computer Science Division \\
University of California, Berkeley \\
\{nemanja, yatish, whitney, kubitron\}@cs.berkeley.edu \\[-0.2in]
} 

\date{}

\begin{document}

\maketitle
\thispagestyle{empty}

\begin{abstract}

We show that the problem of communication in a quantum computer reduces
to constructing reliable quantum channels by distributing high-fidelity
EPR pairs.  We develop analytical models of the latency, bandwidth,
error rate and resource utilization of such channels, and show that
100s of qubits must be distributed to accommodate a single data
communication.  Next, we show that a grid of teleportation nodes forms
a good substrate on which
to distribute EPR pairs.  We also explore the control requirements for
such a network.  Finally, we propose a specific routing architecture and
simulate the communication patterns of the Quantum Fourier Transform to
demonstrate the impact of resource contention.

\end{abstract}

\Section{Introduction}\label{sec:intro}

Quantum computing utilizes properties of subatomic physics to compute in
ways unavailable to classical computers.  As interesting as they may be to
contemplate, quantum computers face a number of barriers to their
implementation, such as the fragility of quantum information and the
lack of systematic techniques for moving such information within the
fabric of a quantum computer.  This paper seeks to address the latter
problem.

It would appear that quantum computers must reach a capacity to process
and store a few thousand quantum bits (or ``qubits'') before becoming
competitive with classical machines.  Since the information contained
within a qubit is extremely fragile, the qubits utilized by algorithms
(called ``logical qubits'') are often implemented by encoding 10s or
100s of physical components (``physical qubits'') using Error Correction
Codes (ECC).  Combine this fact with the need for many temporary qubits
(called ``ancillae'') in quantum algorithms, and we can conclude that an
effective quantum datapath can easily contain a million physical qubits.
We will use ``qubit'' to refer to a physical qubit and ``logical qubit''
explicitly to refer to an encoded state of many qubits.

Anytime such a large number of bits must interact,
communication issues arise: how exactly should
we schedule and route information?
Several observations narrow the space of answers to this question.
First, in all technologies currently under study, two qubits must be
physically adjacent in order to compute a two-input function on these
qubits.  This means that all of the physical qubits comprising one
logical qubit must be moved adjacent to those comprising a second
logical qubit when computing; as a result, each two-input computation
entails movement of 10s or 100s of qubits.  Second, it is not uncommon
for quantum algorithms to require all-to-all communication during some
portion of their execution.  For example, the Quantum Fourier Transform
(QFT)~\cite{Nielsen00a}, a component of Shor's factorization
algorithm~\cite{Shor94},
requires all-to-all communication.  Third, the routing of
qubits must be timed to coincide with the arrival of opcodes to
various functional units.

While a million bits of storage is not particularly large by classical
silicon standards, it \emph{is} large when taking into account the
degradation of state experienced by a qubit through movement.  As
discussed in Section~\ref{sec:fidelity}, for instance, a qubit in an
Ion-Trap computer experiences a probability of corruption of about
$10^{-6}$ when physically transported the distance of a
single storage bit at maximum density. Structuring a million qubits as a
dense 1000$\times$1000 grid, this means that a qubit would experience a
probability of error of more than $10^{-3}$ in traveling from
corner to corner.  Clearly, this is an unacceptable level of error,
leading us to consider other options for moving information.

One solution is to use teleportation~\cite{Oskin03short} in which data is
moved indirectly: after high-fidelity, entangled helper qubits (called
\emph{EPR pairs}) are sent to the endpoints of the desired
communication, they are utilized to transfer the state of a logical
qubit using local quantum operations and reliable classical
communication.  Although EPR pairs experience the same degradation
during movement as data qubits, they represent known states;
consequently multiple lower-fidelity EPR pairs can be combined at the
endpoints to produce high-fidelity EPR pairs through a process called
\emph{purification}.  The process of distributing high-fidelity EPR
pairs to communication endpoints can be viewed as setting up a reliable
``quantum channel'' and is our primary topic.

In the following, we explore architectures for constructing reliable
quantum channels in a large quantum computer.  We start with some
background in Section~\ref{sec:basics}.  In Section~\ref{sec:kontrib},
we discuss architectural options for distributing EPR pairs and
constructing quantum channels.  We show that routing of EPR pairs to
either end of arbitrary points on a quantum computer exhibits much
similarity to routing in classical multiprocessor networks.  Next, in
Section~\ref{sec:fidelity}, we explore the physical resources required
to produce high-fidelity EPR pairs.  One problem that we illuminate is
that the architecture of the network can greatly influence the number of
raw EPR pairs required to set up a communication channel.  We continue
in Section~\ref{sec:simulation} with a simulation of kernels from Shor's
factorization algorithm on a machine that utilizes dimension-ordered
mesh routing to distribute EPR pairs.  Finally, we conclude in
Section~\ref{sec:conclusion}.

\Section{Overview}\label{sec:basics}

In this section, we present an overview of the important aspects of a
quantum computer.  For
concreteness in our analysis, we shall assume the use of ion trap
technology \cite{Kielpinski02}, which has been studied and
demonstrated on a small scale in various experiments
\cite{Leibfried03}.
\SubSection{High-level viewpoint}

\begin{figure}
\begin{center}
\epsfig{file=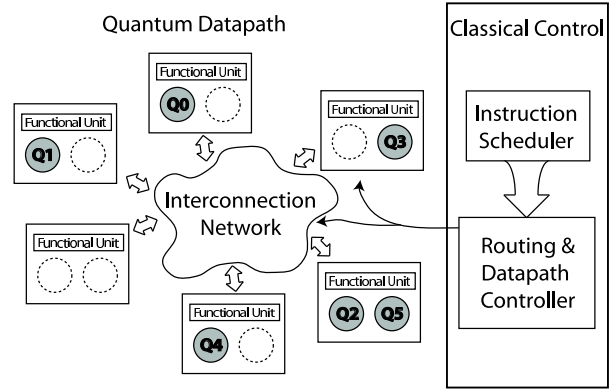,width=0.95\hsize}
\end{center}
\vspace*{-10pt}
\caption{\label{figure:cloud} Abstract view of the quantum datapath.
Each functional unit contains space for two logical qubits, which
may be used for interaction or storage.  A classical instruction
scheduler and datapath control unit initiate communication and
computation.}
\end{figure}

As shown in Figure~\ref{figure:cloud}, our view of a quantum computer
revolves around its quantum datapath.  A set of functional units is
connected through a flexible routing infrastructure.  A classical
control unit transforms a stream of instructions (the quantum algorithm)
into control for both the functional units and the routing
infrastructure.

This figure shows each functional unit operating on one or two logical
qubits.  Functional units must contain registers large enough to hold
their input arguments.  Further, we assume that at least one (and
possibly both) of these registers is capable of holding a logical qubit
for an extended period of time (i.e. capable of continuous error
correction).  Although functional units would appear to contain very
little logic, they are in fact rather large, due to the number of
physical qubits that comprise a single logical qubit\footnote{Note that more
complex functional units are certainly possible, but they are outside the
scope of this paper.}.

We envision routing to be a two-level process.  The classical control
unit schedules communication by specifying a series of logical qubit
movements and functional unit operations.  Classical control logic
within the interconnection network is responsible for efficiently (and
reliably) moving physical qubits as requested by the classical control.

This architecture is justified in the following manner: First, the
number of logical qubits in the system is (relatively) small, leading to
a tractable pairwise scheduling problem for the top-level scheduler.
Second, the number of physical qubits that must be moved in response to
a high level communication request is quite large (especially when
considering raw EPR pairs, as discussed in
Section~\ref{sec:fidelity}). Once a path is constructed from source to
destination, the process of moving these qubits requires relatively
simple (but frequent) control.  Finally, the size of a functional unit
coupled with the need for flexible point-to-point communication leads us
to consider structured routing networks.  An analogy with multiprocessor
networking is very appropriate here, with functional units similar to
compute nodes.

\SubSection{Qubits}

A quantum bit (qubit) is a bit of information encoded in a two-level
quantum system.  The underlying physics of such two-level systems is
potentially quite complicated, but for the purposes of this paper, we
will present a simplified view.  In an ion trap quantum computer, each
qubit is a single ion.  A qubit has some state, which is analogous to a
classical bit's value of 0 or 1, although it can include a
\emph{superposition} or mixture of the two values.  Computation consists
of one-qubit and/or two-qubit gates within functional
units~\cite{Kielpinski02}.

Qubit state is quite fragile.  Current experimental results show the
error rate of a single quantum gate to be around
$10^{-3}$~\cite{Leibfried03}.  Advances in the near future could reduce
this number down to $10^{-6} - 10^{-8}$ \cite{QLA2005metodi,ARDA}, but
it is pretty clear that it will be a long time (if ever) before we reach
error rates found in traditional CMOS gates ($10^{-19}$)
\cite{CMOSerror2000hazucha}.  Even worse, errors can occur during simple
qubit movement, with error probability growing with distance.  These
shortcomings have prompted the development of \emph{quantum} error
correcting codes \cite{Knill97a,Shor95,Steane96} for quantum data.
Quantum ECC codes, much like classical codes, encode a single bit of
data into multiple real bits (ions).

The use of quantum ECC codes leads us to the important distinction
between physical and logical qubits.  A physical qubit is a single
positively charged ion, while a logical qubit is specified as a bit of
data used in the computation.  Physically, a logical qubit is encoded in
some number of physical qubits.  As mentioned previously, we will use
the term ``qubit'' to refer to a physical qubit, while ``logical qubit''
will be stated explicitly.  Logical qubits must be corrected
continuously, before, during and after computation or movement, in order
to combat the ever-present tendency to decohere.  It is not uncommon to
see proposals to use 49 or 343 physical qubits to encode one logical
qubit.

\SubSection{Communication}\label{sec:communication}

Quantum operations involving two logical qubits require the logical
qubits to be physically adjacent.  One of the biggest restrictions to
qubit movement is the \emph{no-cloning theorem}
\cite{nocloning1982wootters} that states that it is impossible to make a
copy of a qubit that is in an arbitrary state.  Consequently, there is
no fan-out in a quantum datapath, and quantum state must be actively
moved to an interaction site before it can participate in a computation.
Thus, qubits stored in non-adjacent parts of our datapath must undergo a
significant amount of movement to perform a two-qubit gate.  In the
following, we discuss two techniques to transfer the state of a qubit
from one point to another.

\paragraph{Ballistic Transport}
An ion trap consists of a set of electrodes which trap an ion in the
space between them.  By placing several ion traps in sequence and
applying specific pulse sequences to the electrodes, we can
ballistically transport the ion along the channel, thus yielding a
simple wire.

\begin{figure}
\begin{center}
\epsfig{file=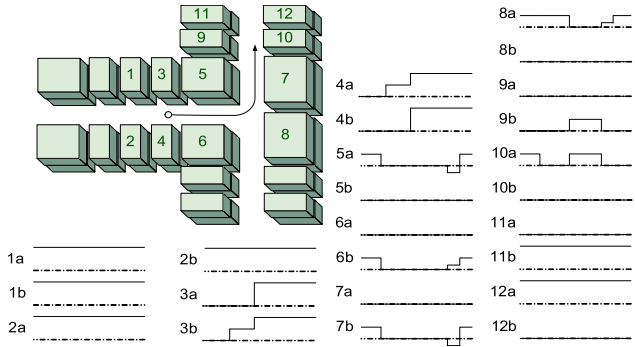,width=\hsize}
\end{center}
\vspace*{-10pt}
\caption{\label{figure:quantum_turn}Electrode layout and waveforms to
move a physical qubit (positively charged ion) from between electrodes 3
and 4 to between 9 and 10.  Gray solids are electrodes; white space
between them is the channel.  The dashed line in each waveform is
ground; a and b refer to top and bottom electrodes at each
location.}
\end{figure}

Figure~\ref{figure:quantum_turn} shows a simplified view of a few ion
traps~\cite{qubitturn05hensinger}, as well as control pulses required to
move an ion through these traps.  There have been demonstrations of MEMS
fabrication techniques that could scale to produce many integrated
qubits~\cite{MEMStrap04madsen}.  In this figure, the gray solids are
electrodes.  The white space between them is the ballistic channel.
Ballistic transport is the most basic communication operation in an
ion-trap computer.  As illustrated by Figure~\ref{figure:quantum_turn},
this seemingly simple operation is relatively complex.

Ballistic movement of a qubit causes some loss in the fidelity of its
state (called decoherence). Thus, there is a limit to the distance that
a qubit may be moved ballistically before error correction must be
performed~\cite{TACO04}.  There is general consensus that any reasonably
sized chip will require an additional form of communication for longer
distances.

\begin{figure}
\begin{center}
\epsfig{file=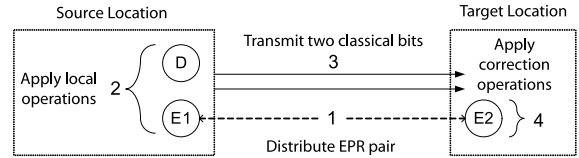,width=0.9\hsize}
\end{center}
\vspace*{-10pt}
\caption{\label{figure:teleport_basic} Teleporting data qubit D to the
target location requires (1) a high-fidelity EPR pair (E1/E2), (2)
local operations at the source, (3) transmission of classical bits, and
(4) correction operations to recreate D from E2 at the target.}
\end{figure}

\paragraph{Teleportation}
Figure~\ref{figure:teleport_basic} gives an abstract view of
teleportation~\cite{Bennett93a}.  We wish to transmit the state of
physical data qubit D from the source location to some distant target
location without physically moving the data qubit (since that would
result in too much decoherence).

We start by interacting a pair of qubits (E1 and E2) to produce a joint
quantum state called an \emph{EPR pair}.  Qubits E1 and E2 are generated
together and then sent to either endpoint.  Next, local operations are
performed at the source location, resulting in two classical bits and
the destruction of the state of qubits D and E1.  Through quantum
entanglement, qubit E2 ends up in one of four transformations of qubit
D's original state.  Once the two classical bits are transmitted to the
destination, local correction operations can transform E2 into an exact
replica of qubit D's original state\footnote{Notice that the no-cloning
theorem is not violated since the state of qubit D is destroyed in the
process of creating E2's state.}. The only non-local operations in teleportation are the
transport of an EPR pair to source and destination and the later
transmission of classical bits from source \emph{to} destination (which
requires a classical communication network).

We can view the delivery of the EPR pair as the process of
\emph{constructing a quantum channel} between source and destination.
This EPR pair must be of high fidelity to perform reliable
communication.  As discussed in Section~\ref{sec:fidelity}, purification
permits a tradeoff between channel setup time and fidelity.  Since EPR
pair distribution can be performed in advance, qubit communication time
can approach the latency of classical communication; of course, channel
setup time grows with distance as well as fidelity.

\begin{figure*}[t]
\centering\epsfig{file=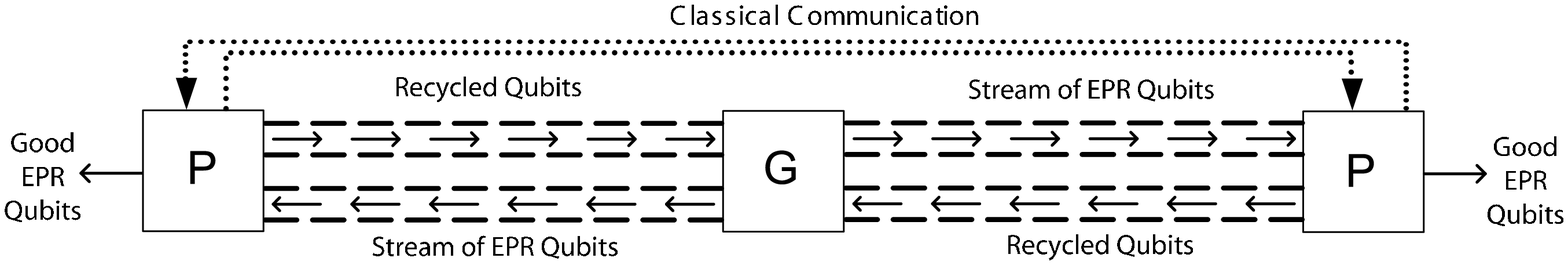,width=0.8\linewidth}
\begin{center}
\begin{minipage}{0.95\linewidth}
\vspace*{-10pt}
\caption{Ballistic Movement Distribution Methodology: EPR pairs are
  generated in the middle and ballistically moved using electrodes.
  After purification, high-fidelity EPR qubits are moved to the logical
  qubits, used, and then recycled into new EPR pairs.
}
\label{figure:saturated_channel2}
\end{minipage}\end{center}
\vspace*{8pt}
\centering\epsfig{file=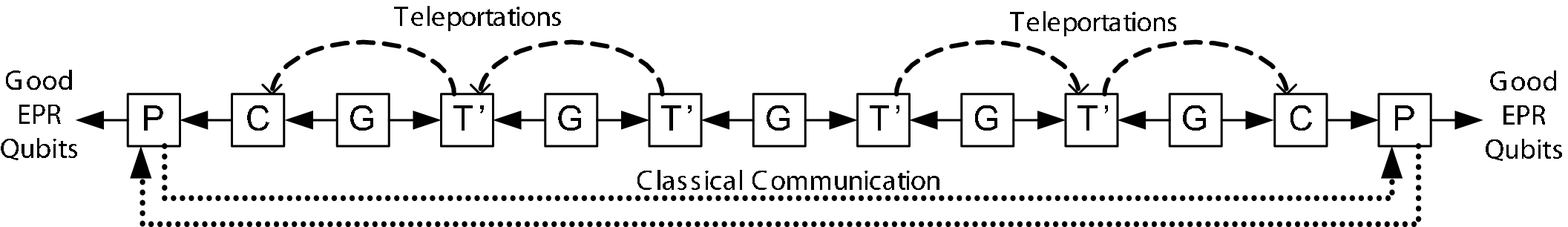,width=0.8\linewidth}
\begin{center}\begin{minipage}{0.95\linewidth}
\vspace*{-10pt}
\caption{Chained Teleportation Distribution Methodology: EPR qubits generated at the
midpoint generator are successively teleported until they reach the
endpoint teleporter nodes before being ballistically moved to corrector nodes
and then purifier nodes.
}
\label{figure:teleport_chaining}
\end{minipage}
\end{center}
\vspace*{-5pt}
\end{figure*}

\Section{Communication Infrastructure}\label{sec:kontrib}

A quantum program execution consists of a sequence of one- and two-qubit
gates applied to a finite set of logical qubits.
The program is run by applying these gates as quickly as possible,
stalling as necessary at each functional unit to wait for logical qubit
operands.  A classical scheduler determines which logical communications
occur and when.  Error correction of logical qubits and fault tolerant
gate implementation are handled at the scheduler level.

As explained in Section~\ref{sec:basics}, ballistic movement of logical
qubits results in far too much decoherence to be practical.  Thus, the
problem of arranging for logical communication comes down to distributing
EPR pairs to the endpoints to allow teleportation of the logical
qubit.





\SubSection{Distribution Methodology}

One option for EPR pair distribution is to generate EPR pairs at
generator (G) nodes in the middle of the path and ballistically
transport them to purifier (P) nodes that are close to the endpoints, as
shown in Figure~\ref{figure:saturated_channel2}.  Purification combines
two EPR pairs to produce a single one of higher fidelity.  For each
qubit in the left purification (P) node, its entangled partner is in the
right P node undergoing the same operations.  For each purification
performed, one classical bit is sent from each end to the opposite one.
Discarded qubits are returned to the generator for reuse.

Another option is to generate an EPR pair and perform a sequence of
teleportation operations to transmit these pairs to their destination.
Correction information from a teleportation (two classical bits) can
be accumulated over multiple teleportations and performed in aggregate at each end of
the chain.  This process is depicted in
Figure~\ref{figure:teleport_chaining}.  A T' node contains units that
perform the local operations to entangle qubits (step 2 in
Figure~\ref{figure:teleport_basic}), but no correction capability (step
4 in Figure~\ref{figure:teleport_basic}).  Instead, each T' node updates
correction information and passes it to the next hop in the chain.

The path consists of alternating G nodes and T' nodes, with a C node and a P
node at each end.  Each G node sends EPR pairs to adjacent T' nodes.  The EPR
pairs generated at the central G node are moved ballistically to the nearest
T' nodes, then successively teleported from T' node to T'
node using the EPR pairs generated by the other G nodes.  Since the
EPR pairs along the length of the path can be pre-distributed, this
method can improve the latency of the distribution if the T' nodes are
spaced far enough apart.


Between each pair of ``adjacent'' T' nodes (as defined by network topology)
is a G node continually generating EPR pairs and sending one qubit of each
pair to each adjacent T' node.  Thus, each T' node is constantly linked with
each neighboring T' node by these incoming streams of entangled EPR qubits.
Each G node is essentially creating a \emph{virtual wire} which connects its
endpoint T' nodes, allowing teleportation between them.

To permit general computation, any functional unit must have a nearby T'
node (although they may be shared).  This implies the necessity of a
grid of T' nodes across the chip, which are linked as described above by
virtual wires.  The exact topology is an implementation choice; one need
not link physically adjacent or even nearby T' nodes, so long as enough
channels are included to allow each G node to be continuously linking
the endpoint T' nodes of its virtual wire with a steady stream of EPR
qubits.  Thus, any routing network could be implemented on this base
grid of T' nodes, such as a butterfly network or a mesh grid.

\SubSection{Structuring Global Communication}\label{sec:comstruct}

As we discussed in Section~\ref{sec:communication}, the process of
moving quantum bits ballistically from point to point presents a
challenging control problem.  Designing control logic to move ions along
a well-defined path appears tractable.  However, controlling every
electrode to select one of many possible paths becomes much more
complex.  Thus, we can benefit from restricting the paths that ions can
take within our quantum computer.  Such a tractable control structure
will involve a sequence of ``single-path'' channels (much like wires in
a classical architecture) connecting router-like control points.

We assume a mesh grid of routers as a reasonable first-cut at a general
purpose routing network \cite{performance94adve,performance90dally}.
Under the Ballistic Movement Distribution Methodology
(Figure~\ref{figure:saturated_channel2}), a routing channel is a
straight sequence of ion traps, while a router is at the intersection.
Under the Chained Teleportation Distribution Methodology
(Figure~\ref{figure:teleport_chaining}), a router is a T' node, and a
routing channel is the pre-generated link between two T' nodes.  In
either case, there must be G nodes distributed across the chip to generate EPR pairs.

\paragraph{Route Planning}
High-level classical control views the quantum datapath at the logical
level.  It tracks the current location of each logical qubit but knows
nothing of the actual encoding used (\emph{i.e.} number of physical
qubits per logical qubit).  This control takes the
sequence of logical operations that comprise the program and identifies
all logical communications that need to occur.  It then begins routing
them while maintaining program order.

Once a path has been determined, EPR pairs need to be generated 
and routed to the endpoints for purification. 
A G node near the middle of the path is given instructions by
the high-level control to generate and name EPR pairs.  These EPR qubits
are then sent from router to router (whether intersection or T' node) along
the routing channels (whether ion traps or teleportation links) until the
endpoints are reached, at which point they are locally routed to the
purifiers.  Thus, under either methodology, the routers need only be able
to make local routing decisions based on a qubit's destination.


\paragraph{Local Routing Control}
Each router and G node needs local classical control to determine how it
handles qubits, which requires a means of identifying qubits.
Thus, each qubit is associated with a classical
message which travels alongside the qubit in a parallel classical network.
The node control for the G node which generates a pair also generates their
accompanying messages.  A qubit's message contains the 
ID assigned by the G node, the destination of this qubit,
the destination of its partner (which is necessary for the
purification steps at the endpoints), and space for the cumulative correction
information that will be used at the endpoint.  
A router forwards a qubit on to the appropriate routing channel or to a local
corrector at the destination.


\begin{figure}
\begin{center}
\centering
\epsfig{file=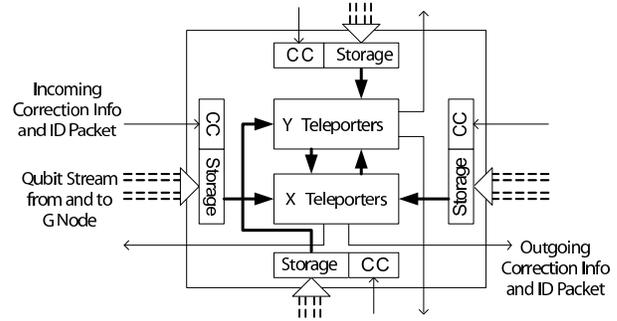,width=0.95\hsize}
\end{center}
\vspace*{-14pt}
\caption{\label{figure:router_2sets2}A Quantum Router: Two sets of
teleporters implement dimension-order routing.  Fat arrows are incoming
qubits from a G node (and recycled ions in opposite direction).  Bold
arrows are ion movement within router.  Thin arrows are classical data.
CC is the classical control including cumulative correction information
and further routing.}
\end{figure}

Figure~\ref{figure:router_2sets2} shows one possible implementation for
a router.
The router receives a constant stream of EPR pairs (from G nodes)
linking it to its neighbor routers.  During an incoming teleportation, a
qubit enters the Storage area to wait for the operations at the
teleportation source to complete.  Classical data in the form of the
teleporting qubit's ID packet and the two classical bits used in the
teleportation enter the adjoining classical control (CC).  The
cumulative correction information is updated in the ID packet, and the
destination information is used to route the qubit to the correct set of
teleporters (or to the nearby C node if this is the endpoint).  For an
outgoing teleportation, a qubit from the G node stream bypasses the
Storage area and moves directly to the appropriate teleporter.


In this router design, the teleporters are divided into two sets.  One
set handles all traffic moving in the X direction, the other handles
traffic moving in the Y direction.  For a turn, an EPR qubit must be
ballistically moved between the teleporter sets (as shown by the
bold-headed arrows).  It is evident from the crossing arrows in the
figure that streams of qubits may need to cross. However, even with
stalling, movement time is so much faster than teleportation
(Table~\ref{table:constants}) that crossing will not be a limiting
factor.




\SubSection{Metrics}\label{sec:metrics}

We shall evaluate various approaches to the EPR distribution mechanism using
six metrics.

\begin{description}
\item[Error Rate:] Ballistic transport and teleportation both cause qubits
to decohere.  The architectural design must take into account the number of
operations each qubit will undergo and the resulting chance for
errors.
\vspace{-8pt}
\item[EPR Pair Count:] While most operations cause qubits to decohere,
purification actually decreases error chance on one pair by sacrificing one
or more other pairs.  The more error that is accumulated, the more pairs
will need to be transported to the endpoints.  
\vspace{-8pt}
\item[Latency:] Logical communication set-up time determines
how far in advance EPR distribution must occur.
\vspace{-8pt}
\item[Quantum Resource Needs:] The quantum datapath resource needs (quantity
of each component) affect chip area and thus
communication distance.
\vspace{-8pt}
\item[Classical Control Complexity:] Generation, ballistic movement,
teleportation and purification must each be controlled classically, so the
classical control requirements vary with communication methodology.
\vspace{-8pt}
\item[Runtime:] Ultimately, we want to know the impact of long distance
communication setup on execution time.
\end{description}

\Section{Qubit Communication Models}\label{sec:fidelity}

In this section, we analyze the communication channels introduced in the
last section.  Before we do this, however, we need to introduce an
important measure called \emph{fidelity} for measuring the difference
between two quantum states.

\SubSection{Fidelity}

\emph{Fidelity} measures the difference between two quantum bit vectors.
Because of quantum entanglement, each of the $2^{n}$ combinations of
bits in a vector of size $n$ are physically separate states.  For a given
problem, one particular vector is considered a reference state that
other vectors are compared against.  For example, if we start with a bit
vector of zeros $[0000]$, and we send the bits through a noisy channel in
which bit 3 is flipped with probability $p$, we would end up with a
probabilistic vector of $((1-p)[0000] + p[0010])$.  The fidelity of the
final state in relation to the starting (``error-free'') state is just
$1-p$.  So, in the case of an operational state vs. a reference
"correct" state, the fidelity describes the amount of error introduced
by the system on the operational state~\cite{Nielsen00a}.  A fidelity of
1 indicates that the system is definitely in the reference state.  A
fidelity of 0 indicates that the system has no overlap with the
reference state.

We characterize errors by calculating the fidelity of qubits traversing
the various quantum channels and gates necessary to route and move bits
around a communication network.  We will combine models of the
individual communication components so that we get an overall
\emph{fidelity of communication} as a function of distance and architecture.


\SubSection{Ion Trap Parameters}

For the remainder of the paper, we utilize parameters for ion trap
quantum computers.  We use the experimental values for time constants
shown in Table~\ref{table:constants}
\cite{QLA2005metodi,Ozeri05,Rowe02}.  A ``cell'' refers to the minimum
distance of a ballistic move (one ion trap).  Both teleportation and
purification require classical bits to be routed between the endpoints,
and thus both of these numbers vary with distance.  Further, every
quantum operation other than purification results in errors (loss of fidelity).
Table~\ref{table:probconstants} lists the error probabilities used for
the following fidelity calculations.

\SubSection{Ballistic Transport Model}

In ballistic movement, the fidelity of a bit after going through the
ballistic channel over $D$ cells is:
\begin{eqnarray} F_{new} = F_{old}(1-p_{mv})^D \end{eqnarray}
since each hop introduces a probability of error.
The time to
perform ballistic movement is given in time per cell moved through and
from Table~\ref{table:constants} is $0.2\mu s/cell$.
\begin{eqnarray} \label{eqn:ballistic_lat} t_{ballistic} = t_{mv} \times D
\end{eqnarray}

\begin{table}
\centering
\begin{tabular}{|l|c|r|}
\hline
Operation & Variable Name & Time ($\mu$s) \\
\hline \hline
One-Qubit Gate & $t_{1q}$ & 1 \\
\hline
Two-Qubit Gate & $t_{2q}$ & 20 \\
\hline
Move One Cell & $t_{mv}$ & 0.2 \\
\hline
Measure & $t_{ms}$ & 100 \\
\hline
Generate & $t_{gen}$ & 122 \\
\hline
Teleport & $t_{tprt}$ & $\sim$122 \\
\hline
Purify & $t_{prfy}$ & $\sim$121 \\
\hline
\end{tabular}
\caption{ Time constants for operations in ion trap technology.  One
cell is the minimum distance for ballistic movement (one ion trap).}
\label{table:constants}
\vspace*{10pt}
\centering
\begin{tabular}{|l|c|r|}
\hline
Operation & Variable Name & Error Probability \\
\hline \hline
One-Qubit Gate & $p_{1q}$ & $10^{-8}$ \\
\hline
Two-Qubit Gate & $p_{2q}$ & $10^{-7}$ \\
\hline
Move One Cell & $p_{mv}$ & $10^{-6}$ \\
\hline
Measure & $p_{ms}$ & $10^{-8}$ \\
\hline
\end{tabular}
\caption{Error probability constants for various operations in ion trap
technology.  Estimates come from \cite{QLA2005metodi, ARDA}.}
\label{table:probconstants}
\end{table}

\SubSection{Teleportation Transport Model}

The fidelity of a qubit teleportation is more complicated because it
involves a combination of single and double qubit gates ($p_{1q}, p_{2q}$) and
qubit measurement ($p_{ms}$) \cite{Dur98a}:
 \begin{eqnarray} 
 F_{new}&=&\frac{1}{4}\left(1+3(1-p_{1q})(1-p_{2q}) \frac{(4(1-p_{ms})^2 - 1)}{3}\right. \nonumber\\
 	&\times&\left. \frac{(4 F_{old}-1)(4 F_{EPR}-1)}{9}\right)\label{eqn:teleport}
 		\end{eqnarray}
The fidelity after a teleportation involves the fidelity of the data
before teleportation ($F_{old}$) and the fidelity of the EPR pair used
to perform the teleportation ($F_{EPR}$).

Although ballistic movement error does not appear in this formula, it
should be mentioned that the fidelity of the EPR pair will be degraded while
being distributed to the endpoints of the teleportation channel.  Thus,
even though the qubit undergoing teleportation incurs no error from
direct ballistic movement, there is still fidelity degradation due to
EPR pair distribution.

We produce EPR pairs from two qubits initialized to the zero state
using a few single and double qubit gates.  The fidelity of an EPR pair
immediately after generation is:
\begin{eqnarray} F_{gen} \propto (1-p_{1q})(1-p_{2q})F_{zero} \end{eqnarray}
$F_{zero}$ is the fidelity of the starting zeroed qubits.
Generation time involves one single and one double qubit gate.  As mentioned
in Table~\ref{table:constants}, this time is projected to be $21\mu s$.

If we assume that EPR pairs are already located at the endpoints of our
channel, teleportation time is given in Table~\ref{table:constants} as
$122\mu s$ and has the form:
\begin{eqnarray} \label{eqn:teleport_lat} t_{teleport} = 2t_{1q} +
t_{2q} + t_{ms} + t_{classical\ bit\ mv} \times D \end{eqnarray}

\SubSection{\label{sec:EPRpurify}EPR Purification Model}\label{sec:epr_purif}


As shown by Equation~\ref{eqn:teleport}, the fidelity of the EPR pairs
utilized in teleportation ($F_{EPR}$) has a direct impact on the
fidelity of information transmitted through the teleportation channel.
Since EPR pairs accrue errors during ballistic movement, teleportation by
itself is not an improvement over direct ballistic movement of data
qubits unless some method can be utilized to improve the fidelity of EPR
pairs.

\begin{figure}[t]
\begin{center}
\epsfig{file=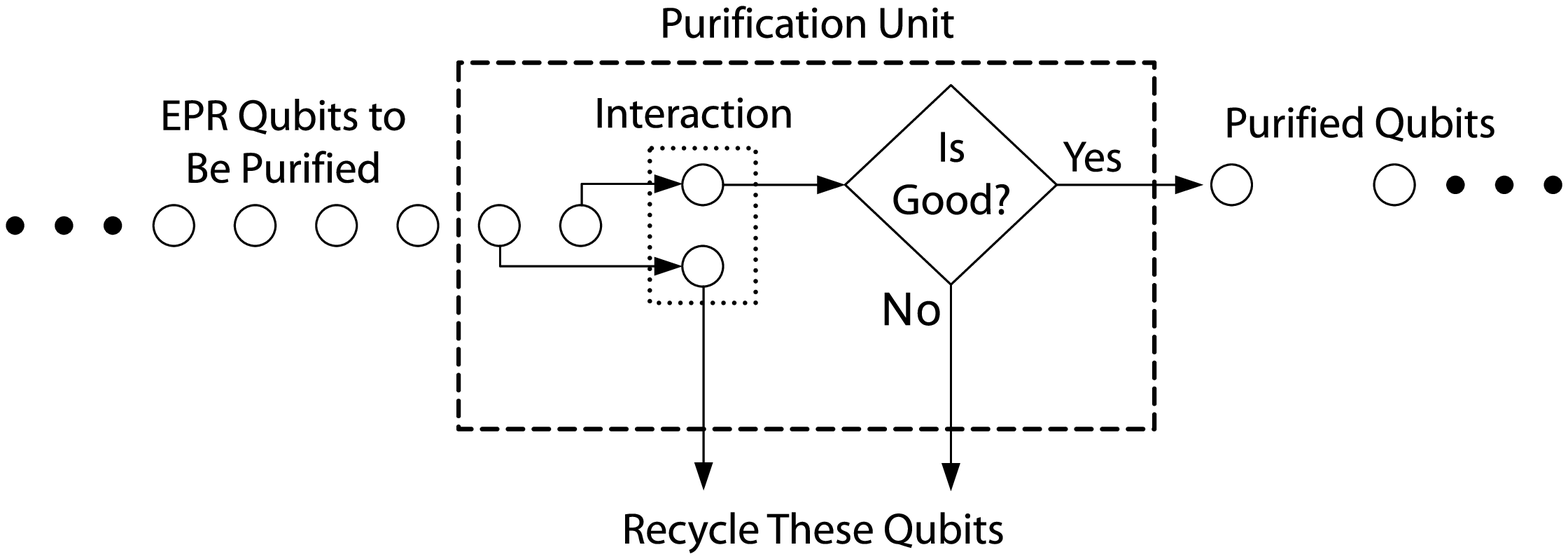,width=0.9\hsize}
\end{center}
\vspace*{-16pt}
\caption{Simple Purification: pairs of EPR qubits undergo local
  operations, yielding a classical bit that is exchanged with the
  partner unit.  Purification succeeds if classical bits are equal.}
\label{figure:purify_basic}
\vspace{5pt}
\begin{minipage}[t]{3.3in}
\begin{center}
\epsfig{file=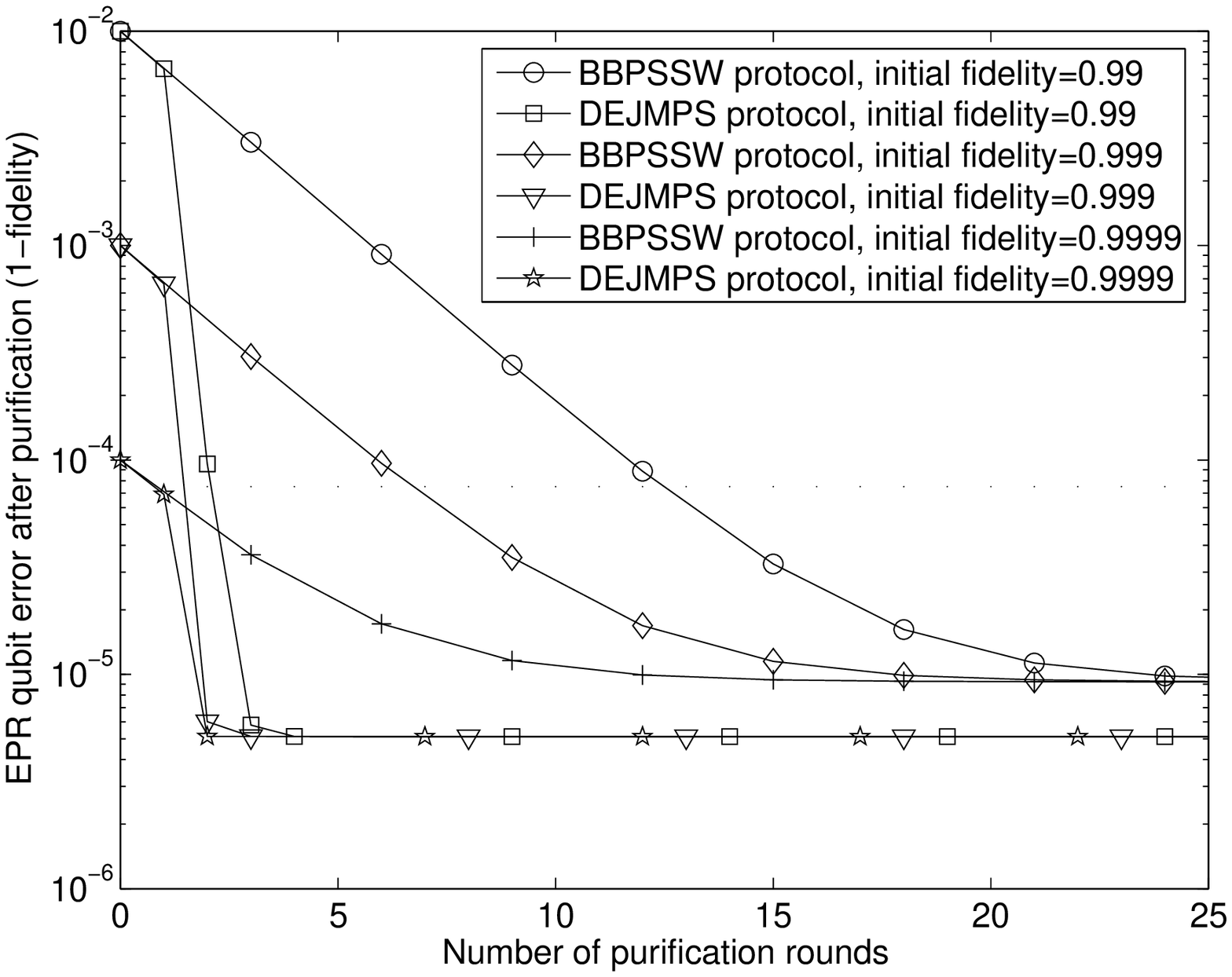,width=.9\hsize}
\end{center}
\vspace*{-14pt}
\caption{\label{fig:purify_rounds_pairsAB}Error rate (1-fidelity) for
  surviving EPR pairs as a function of the number of purification rounds
  (tree levels) performed by the DEJMPS or BBPSSW protocol.  Lower is
  better.}
\end{minipage}
\end{figure}

Purification combines two lower-fidelity EPR pairs with local operations
at either endpoint to produce one pair of higher fidelity; the remaining
pair is discarded after being measured.
Figure~\ref{figure:purify_basic} illustrates this process, which must be
synchronized between the two endpoints since classical information is
exchanged between them.  On occasion both qubits will be discarded
(with low probability).

The purification process can be repeated in a tree structure to obtain
higher fidelity EPR pairs.  Each {\it round} of purification corresponds
to a level of the tree in which all EPR pairs have the same fidelity.
Since one round consumes slightly more than half of the remaining pairs,
resource usage is exponential in the number of rounds.  There are two
similar tree purification protocols, the DEJMPS
protocol~\cite{purify96deutsch} and the BBPSSW
protocol~\cite{Bennett96b}.  The analysis of the DEJMPS protocol
provides tighter bounds which assures faster, higher fidelity-producing
operation compared to the BBPSSW protocol.
The effects are significant, implying that purification mechanisms must
be considered carefully\footnote{Dur also proposes a linear approach to
purification~\cite{Dur98a}; unfortunately, it appears to be sensitive to
the error profile. We will not analyze it here.}.

Figure \ref{fig:purify_rounds_pairsAB} shows error rate as
a function of number of purification rounds.  The BBPSSW protocol takes
5-10 times more rounds to converge to its maximum value as the DEJMPS
protocol.  Since EPR pair consumption is exponential in number of
rounds, the purification protocol has a
large impact on total EPR resources needed for communication.  Other
features of Figure \ref{fig:purify_rounds_pairsAB} to note are that
DEJMPS has higher maximum fidelity and converges to maximum fidelity
faster than BBPSSW (possibly because BBPSSW partially randomizes its
state after every round).



Finally, the time to purify a set of EPR qubits is dependent on the
initial and desired fidelity.  The time to complete one round of
purification is $121\mu s$ from Table~\ref{table:constants}:
\begin{eqnarray} t_{purify\ round} = t_{2q} + t_{ms} + t_{classical\
bit} \end{eqnarray}

\SubSection{Communication Model Analysis}

We know from the most recent version of the threshold
theorem for fault-tolerant quantum computation that data qubit fidelity
must be maintained above $1-7.5*10^{-5}$~\cite{localfault04svore}.
Because the preservation of data qubit fidelity is our highest priority,
we choose to transport all data by way of single teleports, since this
introduces the minimum error from ballistic movement.  Additionally, to
minimize the number of interactions with expendable, lower fidelity EPR
pairs, we choose to move a data qubit with a single, long-distance
teleportation.  This necessitates the distribution of EPR pair qubits to
communication endpoints.  Since data qubits interact with these EPR
pairs, the above threshold must be imposed on them to avoid tainting the data.


Two options present themselves for distributing high-quality EPR
pairs to channel endpoints.  First, one could ballistically move the EPR
pairs to the endpoints, which is preferable to moving data ballistically
because EPR pairs can be sacrificed if they accumulate too much error.
Second, one could route EPR pairs through a series of teleporters, as
shown in Figure~\ref{figure:teleport_chaining}.  While preserving
fidelity of our data states is top priority, when dealing with less
precious EPR pairs, we do not have to adhere to strict maximal fidelity
preserving distribution methods.  In the rest of this section, we will
investigate the tradeoffs between ballistic distribution and chained
teleportation distribution of EPR pairs.

\paragraph{Fidelity Difference} The final fidelity of these two
techniques is approximately the same. Conceptually, the final EPR pair
either directly accumulates movement error (through ballistic movement)
or is interacted with several other EPR pairs to teleport it to the
endpoints and these intermediate EPR pairs have accumulated the same
distance ballistically.  By interacting with intermediate pairs, the
final pair accumulates all this error.  This statement assumes that the
fidelity loss from gate error is much less than the loss due to
ballistic movement, which is the case for ion traps, as shown in
Table~\ref{table:constants} (for two teleporters spaced 100 cells apart,
ballistic movement error equals $1-(1-10^{-6})^{100} \approx 10^{-4}$
compared to $10^{-7}$ for a two-qubit gate error).

\begin{figure}[t]
\begin{minipage}[t]{3.3in}
\epsfig{file=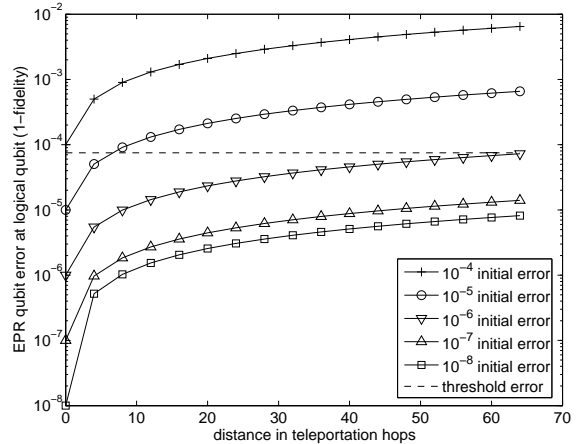,width=.9\hsize}
\vspace*{-12pt}
\caption{\label{fig:sensitivity_initial_final}Final EPR error
  (1-fidelity) as a function of number of teleportations performed, for
  various initial EPR fidelities.  The horizontal line represents the
  minimum fidelity the EPR pair must be at to be suitable for
  teleportation of data qubits, $1-7.5*10^{-5}$}
\end{minipage}
\end{figure}

Long-distance distribution of EPR pairs can severely reduce the fidelity
of the EPR pairs arriving at a functional unit for data teleportation,
as shown in Figure~\ref{fig:sensitivity_initial_final}.  In order to
process 1024 qubits, we could imagine arranging them on a square 32x32
grid, in which the longest possible Manhattan distance is 64 logical
qubit lengths.  If we assume that we have teleporter units at every
logical qubit, EPR pair distribution could require up to 64 teleports.
From the figure, teleporting 64 times could increase EPR pair qubit
error by a factor of 100.  The dotted line represents the threshold at
which the EPR pairs must be in order to not corrupt the data qubit
when teleporting it.  In order to preserve data fidelity, we must use
EPR pair purification.  One way to think about this process is to stitch
Figures~\ref{fig:purify_rounds_pairsAB}
and~\ref{fig:sensitivity_initial_final} side-by-side, so that EPR pairs
accumulate error (degrade in fidelity) as they are teleported and then
purified back to a higher fidelity at the endpoints before being used
with data.


\paragraph{Latency Difference} Equation~\ref{eqn:ballistic_lat}
shows a linear dependence on distance for ballistic movement latency.
Equation~\ref{eqn:teleport_lat} also shows that teleportation has a
linear dependency on distance as well, but the constant in this case is
for the necessary classical communication.  We assume classical
information can be transferred on a time scale orders of magnitude
faster than the quantum operations.

If teleportation is considered performed in near constant time, then we
would like to know the distance crossover point where teleportation
becomes faster than the equivalent ballistic transport.  From
Table~\ref{table:constants}, teleportation takes about $122\mu s$ while
ballistic movement takes $0.2 \mu s$ per ion trap cell.  Thus for a
distance of about 600 cells, teleportation is faster than ballistic
movement.
We assume our communications fabric to be a 2-D mesh of teleporter
nodes and use 600 cells as the distance that each teleportation
``hop'' travels.  Allowing teleportations of longer distances would
further reduce communication latency in some cases but would then
require more local ballistic movement to get an EPR pair from the
nearest teleporter to its final destination.


\SubSection{Purification Resources}

Earlier in this section, we noted that when we purify a set of EPR
pairs, we measure and discard at least half of them for every iteration.
This means that to perform $x$ rounds, we need more than $2^x$ EPR pairs
to produce a single good pair.

To measure EPR resource usage, we count the total number of pairs used
\emph{over time} to move a level 2~\cite{Steane96} error
corrected logical data qubit between endpoints.  This means we are
transporting 49 physical data
qubits some distance by way of teleportation.  We find that the
total number of EPR qubits necessary to move a datum
critically affects the data bandwidth that our network can support.
This metric differs from that used in a number of proposals for quantum
repeaters which focus on the layout of a quantum teleporter and are most
concerned with spatial EPR resources, i.e. how much buffering is
necessary for a particular teleporter in the
network~\cite{opticalrepeater05childress}.  We will show that our design
is fully pipelined, and therefore only a small number of qubits must be
stored at any place in the network at any time.

We saw in Figure~\ref{fig:purify_rounds_pairsAB} that if we start at a
relatively low fidelity and try to obtain a relatively high fidelity,
we could need more than a million EPR pairs to produce a single high
fidelity pair using the BBPSSW protocol.  Therefore we use the DEJMPS
protocol in all further analysis.  Even though the DEJMPS protocol
converges to good fidelity values much quicker, the exponential
increase in resources for each additional round performed means we
must be careful about how much error we accumulate when distributing 
EPR pairs.  We will also show that the point in the datapath at which
purification is performed can have a dramatic impact on total EPR
pairs consumed.  We have 3 options:

\begin{description}
\item[Endpoints only:]Purify only at the endpoints, immediately before using
EPR pairs to teleport data.
\vspace{-8pt}
\item[Virtual wire:] Purify EPR pairs which
create the links between teleporters, namely the constant stream of
pairs from a G node to adjacent T' nodes. The result is higher fidelity qubits used for chained teleportation.
\vspace{-8pt}
\item[Between teleports:]Purify EPR pairs \emph{after} every
teleportation; this purifies qubits that are being chain
teleported rather than  qubits assisting the chained teleportation.
\end{description}

\begin{figure}[t]
\begin{center}
\epsfig{file=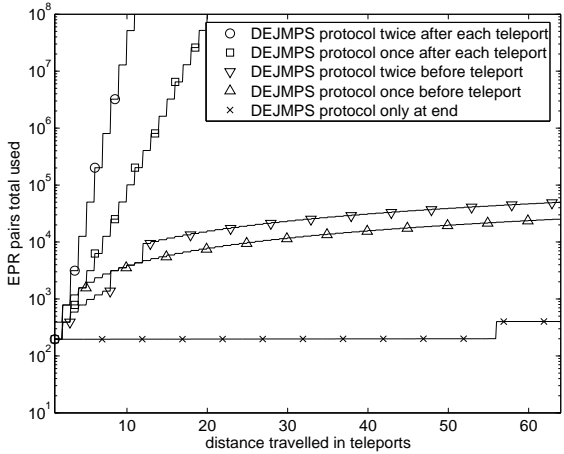,width=.9\hsize}
\end{center}
\vspace{-14pt}
\caption{\label{fig:purifyscheme_localB}Total EPR pairs
consumed as a function of distance and point at which
purification scheme DEJMPS is performed.}
\end{figure}

We now model the error present in our
entire communication path.  Assuming the EPR pairs at the logical qubit
endpoints must be of fidelity above threshold, we determine the number
of EPR pairs needed to move through different parts of the network per
logical qubit communication.

\paragraph{Total EPR Resources} 
Figure~\ref{fig:purifyscheme_localB} shows that the Endpoints Only
scheme uses the fewest total EPR resources.  This conclusion is evident
if we refer back to Figure~\ref{fig:purify_rounds_pairsAB}, where
purification efficiency asymptotes at high fidelity; thus,
purifying EPR pairs of lower fidelity shows a larger percentage gain in
fidelity than purifying EPR pairs of high fidelity.  From this, we can
see that to minimize total EPR pairs used in the whole system, it makes
sense to correct all the fidelity degradation in one shot, just before
use.

\paragraph{Non-local EPR Pairs}
Another metric of interest is to focus only on those EPR pairs that are
transmitted to endpoints during channel setup (i.e. those that are
teleported through the path).  This resource usage is critical for
several reasons: First, every EPR pair moved through the network
consumes the slow and potentially scarce resource of teleporters; in
contrast, the EPR pairs consumed in the process of producing virtual
wires are purely local and thus less costly.  Second, because of
contention in the network, EPR pairs communicated over longer distances
(multiple hops) place a greater strain on the network than those that
are transmitted only one hop.  The channel setup process can be
considered to consume bandwidth on every virtual wire that it traverses.
Third, the total EPR pairs
transmitted to endpoints during channel setup consumes purification
resources at the endpoints---a potentially slow, serial process.

\begin{figure}[t]
\begin{center}
\epsfig{file=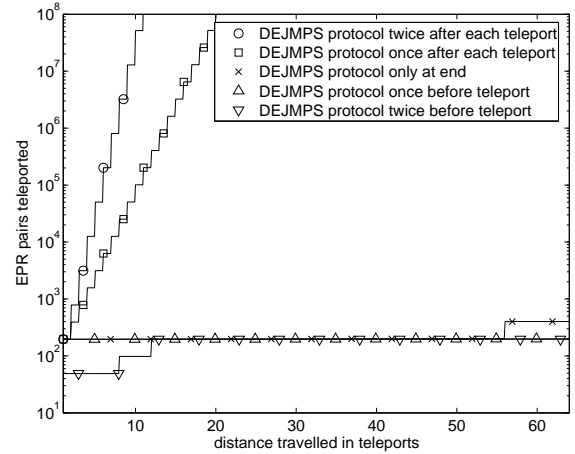,width=.9\hsize}
\end{center}
\vspace{-14pt}
\caption{\label{fig:purifyscheme_nonlocalB}Total EPR pairs in
teleportation channel as a function of distance and point in transport
in which purification scheme DEJMPS is performed.  The only 2 lines that change from
Figure~\ref{fig:purifyscheme_localB} are the purify before teleport cases.}
\end{figure}

Figure~\ref{fig:purifyscheme_nonlocalB} shows that purifying EPR pairs
after each teleport transmits many more EPR pairs than purifying at the
endpoints (either with or without purifying the virtual wires).  From
this figure, we see that over-purifying bits leads to additional
exponential resource requirements without providing improved final EPR
fidelity\footnote{The authors of \cite{opticalrepeater05childress} claim
that this nested purification technique (after every teleport) has small
resource requirements; however, they count spacial resources rather than
total resources over time.}.  Virtual wire purification improves the
underlying channel fidelity for everything moving through the
teleporters, thereby allowing less error to be introduced into qubits
traveling through the channel.  For a given target fidelity at the
endpoints, virtual wire purification reduces the number of EPR pairs
that need to move through the teleporters and also reduces the
strain on the endpoint purifiers.



To summarize, we have made the following design decisions based on
fidelity and latency concerns:
\begin{description}
\item[Teleport data always:]  Data qubits sent to destination with 
  single teleportation to minimize ballistic error.
\vspace{-8pt}
\item[Teleport EPR pairs:] EPR pairs distributed to endpoints with
  teleportation, allowing pre-purification to increase the overall
  fidelity of the network.

\vspace{-8pt}
\item[Purification before teleport and at endpoints:] Purify
intermediate EPR pairs before they are used for teleportation as well as
EPR pairs at the channel endpoints.
\end{description}

\begin{figure}[t]
\epsfig{file=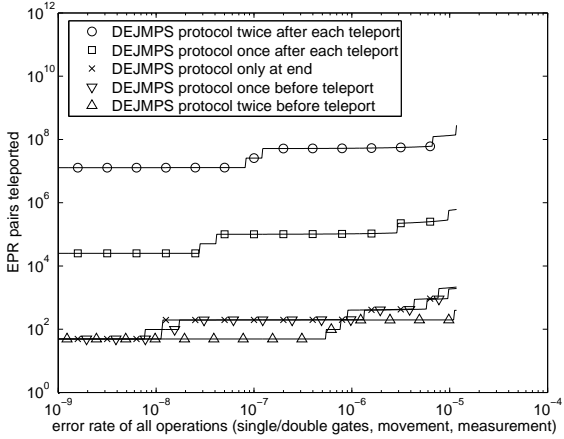,width=.9\hsize}
\vspace*{-8pt}
\caption{\label{fig:sensitivity_error_rateB}Number of EPR pairs that
  need to be teleported to support a data communication within the
  error threshold.  All error rates are set to the rate specified on
  the x-axis.}

\end{figure}
Finally, Figure~\ref{fig:sensitivity_error_rateB} shows the sensitivity
of the EPR resources necessary to sustain our previous error threshold
goals as a function of the error of the individual operations like
quantum gates, ballistic movement, and quantum measurement.  The first
thing to note are the abrupt ends of all the plots near $10^{-5}$.  This
is the point at which our whole distribution network breaks down, and
purification can no longer give us EPR pairs that are of suitably high
fidelity (above $1-7.5*10^{-5}$).  The fact that all the purification
configurations stop working for the same error rate is due to the fact
that the purification schemes we investigated are limited in maximum
achievable fidelity by operation error rate and not the fidelity of
incoming EPR pairs (unless the fidelity is {\it really} bad).
Throughout the regime at which our system {\it does} work however, the
total network resources only differ by a factor of up to 100 for a
10,000 times difference in operation error rate.

\Section{Simulation}\label{sec:simulation}

\begin{figure}
\begin{center}
\epsfig{file=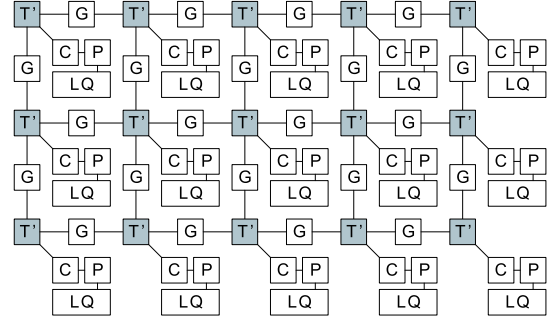,width=0.85\hsize}
\end{center}
\vspace*{-12pt}
\caption{\label{figure:sim_layout5}Sample Layout of a 5x3 mesh grid containing
Logical Qubits (LQ) and G, T', C and P nodes (not to scale).}
\end{figure}

We built an event-driven communication simulator using Java to explore the
effects of parameter variation (number of generators, teleporters and purifiers)
and resource contention on the runtime of a realistic execution.
The simulator accepts an instruction stream consisting of a sequence
of two-logical-qubit operations 
and a layout
of the communication grid
constructed of the following units: Teleporters, Purifiers, Generators, Logical
Qubits, and Wires.  
Simulations were performed on a
16x16 grid of logical qubits, using a mesh grid interconnect topology,
depicted in part in Figure~\ref{figure:sim_layout5} and using
operation latencies shown in Table~\ref{table:constants}. 

The logical instruction stream is processed by a control unit which
determines a path for each logical communication and creates the necessary
control messages.
The scheduler attempts to execute as many logical instructions in
parallel as possible while maintaining instruction order
dependencies, using dimension order routing to generate paths.
Improving this component is a topic for future research.  Teleporters
in each T' node are partitioned into
two equal sets, as shown in Figure~\ref{figure:router_2sets2}, with
each set time multiplexed to prevent blocking of channels that share
T' nodes.

We consider two different architectures based on the topology in
Figure~\ref{figure:sim_layout5}.  One approach is to define each
LQ node as a Home Base for a single logical qubit, with the capability to
error correct that logical qubit (including room for all necessary ancillae
and local communication) and with enough room to allow another logical qubit
to teleport in for a two-logical-qubit operation, requiring each
logical qubit to teleport home after each logical operation.


Another possibility is for each LQ node to have room to error correct two
logical qubits.  This eliminates the need to teleport home but increases
the size of each LQ node.  This architecture shall be referred to as the
Mobile Qubit layout.

\SubSection{Purifier Implementation}\label{sec:purify_imp}

\begin{figure}
\begin{center}
\epsfig{file=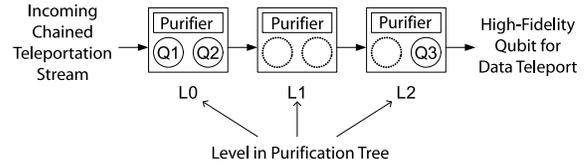,width=0.9\hsize}
\end{center}
\vspace*{-15pt}
\caption{\label{figure:queue_purifier2} Robust
tree-based purification: Incoming qubits are purified
once at L0, representing the lowest level of the purification
tree.  Successfully purified qubits are moved on to L1 and purified there,
representing the second lowest level, and so on.
}
\end{figure}


We could implement tree purification (Section~\ref{sec:epr_purif})
naively at each possible endpoint by including one hardware purifier for
each node in the tree.  However, as the tree depth increases, the
hardware needs quickly become prohibitive.  Additionally, this mechanism
provides no natural means of recovering from a failed purification (loss
of a subtree).


A more robust queue-based purifier implementation is shown in
Figure~\ref{figure:queue_purifier2}.  
There are three advantages of this implementation.  First,
a tree structure of depth $n$ is implemented with $n$ purifiers (rather
than $2^n-1$, as above).  Second,
movement between levels of purification is minimized, lessening the
impact of movement (which is over an order of magnitude worse
than two-qubit gate error; see Table~\ref{table:constants}).  Third,
no special handling for lost subtrees due to failed purifications 
is necessary as they'll be rebuilt naturally.

The primary drawback of this implementation is the latency penalty.
If $x$ purifications are needed at level L0, then they must
necessarily be done sequentially.  This problem may be alleviated
by including more queues, however, since each logical communication
requires multiple high-fidelity EPR pairs, depending upon the encoding
used.  For these reasons, we use Queue Purifiers in our simulations.

\begin{figure}
\begin{center}
\epsfig{file=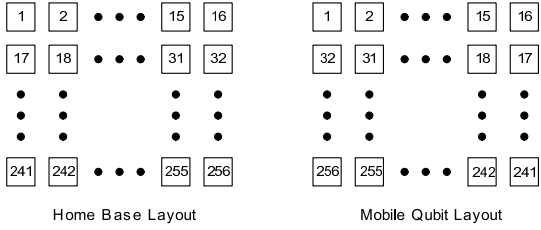,width=.85\hsize}
\end{center}
\vspace*{-12pt}
\caption{\label{figure:LQ_grids}Two possible logical qubit layouts.  The
Mobile Qubit Layout capitalizes on the sequential nature of QFT.}
\end{figure}

\SubSection{Benchmarks}
We studied Shor's Factorization Algorithm \cite{Shor94}, which consists
of three communication-intensive components: Quantum Fourier Transform
(QFT) on one set of logical qubits, Modular Exponentiation (ME) on
another set, and a Modular Multiplication (MM) between the two sets.

QFT contains all-to-all communication between the logical qubits.  MM has
a bipartite communication pattern, with all from one set
communicating with all from the other set.  ME consists of squaring
steps which require all-to-all communication and multiplication steps which
involve bipartite communication.  This provides us with two benchmark
communication patterns.  Since QFT is also a component of many other quantum
algorithms \cite{Ip02,Jozsa98,Jozsa00,Kitaev95,Mosca99}, we
decided to concentrate study on the all-to-all pattern.

A circuit for performing QFT is described in \cite{Nielsen00a}.
Given $n$ logical qubits, labeled 1, 2,... $n$, each logical qubit must
interact once with each other logical qubit, in
numerical order.  Thus, the first few communications in QFT are 1-2, 1-3,
(1-4, 2-3), (1-5, 2-4), (1-6, 2-5, 3-4), where communications in parentheses
may occur simultaneously.

When simulating the Home Base implementation described earlier, we
utilize the basic layout shown on the left of
Figure~\ref{figure:LQ_grids}.  Since QFT is a common kernel, it's
worthwhile to optimize a bit.  So in the case of Mobile Logical Qubits,
we simulate the Mobile Qubit Layout in Figure~\ref{figure:LQ_grids}.  In
this layout, logical qubit 1 successively teleports from logical qubit
to logical qubit, being error corrected in place after each logical
operation.  Once logical qubit 1 has passed, logical qubit 2 can start
moving along the line, and so on.  Once a logical qubit has completed
its interaction with logical qubit 256, it is teleported back to its
starting location.  Thus, this particular circuit consists primarily of
local communications, with the exception of teleports from the final
logical qubit location.

\begin{figure}[t]
\begin{center}
\epsfig{file=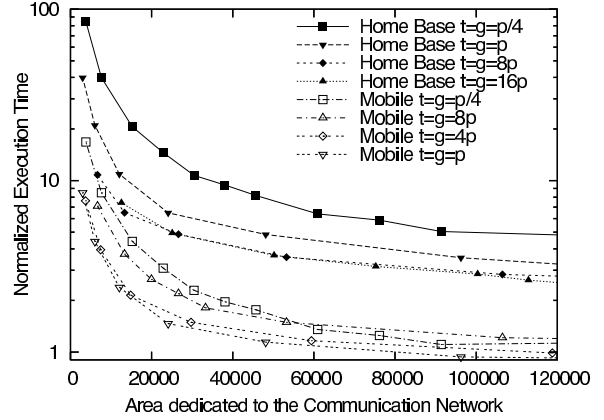,width=0.95\hsize}
\end{center}
\vspace*{-14pt}
\caption{\label{figure:combined-area} Benchmark execution normalized to
execution on $t=g=p=1024$, as a function of resource allocation.  }
\vspace*{-3pt}
\end{figure}

\SubSection{Results}
We studied the effects of resource allocation on execution time.  Since
generation and teleportation have nearly equivalent latency, we match
their bandwidth by setting the number of generators $g$ per G node to
equal the number of teleporters $t$ per T' node.  To avoid deadlock, storage
for incoming teleports is not multiplexed, 
yielding $t$ storage cells per incoming link (yielding $4t$
storage cells per T' node).  Given results in Section~\ref{sec:fidelity},
we will need a maximum purification tree of depth three (for distances
under consideration); consequently, we use Queue Purifiers of depth
three.

Figure~\ref{figure:combined-area} shows 
 the execution time of the benchmarks.  Since the expected number
of EPR pairs required for the longest communication path is $392$
($=$ pairs for endpoint purification $\times$ qubits per logical qubit
$= 2^3 \times 49$), we
normalized execution time to the execution time for $t=p=g=1024$ as a
close approximation of unlimited resources.

By fixing the area dedicated to the interconnection network (T', G,
and P nodes) and varying the size of T' and G nodes relative to
P nodes, we can demonstrate where the bottlenecks in the
system arise.  The Home Base benchmark contains many simultaneously
active channels sharing the T' nodes.  As more channels share T'
nodes, the time multiplexing limits the overall bandwidth of each
channel, minimizing the number of purifiers necessary at the end points.
As shown in the graph, the limited bandwidth of the channels allows us
to allocate more resources to the T' nodes by sacrificing the number
of Queue Purifiers.

In contrast, the Mobile Qubit benchmark contains fewer channels
sharing T' nodes, placing a higher demand on the number of Queue Purifiers
available at the end point.  As we dedicate more and more of the
available resources away from P nodes to T' nodes, the performance
suffers, as shown in the difference between $t=g=4p$ and $t=g=8p$.












\Section{Conclusion}\label{sec:conclusion}
In this paper, we explored designs for interconnection networks of
quantum computers.  We accounted for both the flow of quantum bits and
the classical information that accompanies it.  We simulated a mesh grid
architecture under the communication patterns of a common component of
many quantum algorithms, the Quantum Fourier Transform, to determine the
effects resource allocation will have on performance.  We show that
devoting sufficient resources to the network is important for
performance.

Our study revealed how qubit fidelity is dependent on the errors in 
quantum operations and the distance of communication.  Fidelity
degradation is also strongly dependent on the choice of the purification
algorithm.  Even under the most optimal circumstances, the number of
EPR pairs distributed to set up a communication channel is several
dozen.  This implies the need for an EPR pair distribution
infrastructure in a quantum datapath.  Not only does this impose the
allocation of space for active components (such as teleporters and
generators), but it also necessitates temporary storage for qubits as well
as an efficient recycling mechanism to allow the constant reuse of
qubits.

Finally, we highlighted the need for a classical network to
organize this infrastructure.
The network must have adequate bandwidth for one in-flight message for
each physical qubit in the system as well as the classical bits for each
teleportation and purification operation.

\renewcommand{\refname}{References\vspace*{-9pt}}
\bibliographystyle{latex8}
\bibliography{main,qubib,micro38-QuantumQLA-paper}
\end{document}